\documentclass[aps,twocolumn,groupedaddress,amsmath,amssymb]{revtex4}

\usepackage{graphicx}
\usepackage{dcolumn}
\usepackage{xcolor}%
\usepackage{booktabs}%
\usepackage{bm}
\usepackage{color, soul}
\usepackage{amsmath}
\usepackage{amssymb}

\begin{document}

\title{One-step synthesis of Cu-doped Pb$_{10}$(PO$_{4}$)$_{6}$Cl$_{2}$ apatite: A wide-gap semiconductor}

\date{\today}

\author{Wuzhang Yang$^{1,2,3}$\footnote{These authors contributed equally to this work.}}
\author{Zhihong Pang$^{1\ast}$}
\author{Zhi Ren$^{1,2}$\footnote{Electronic Supplementary Information (ESI) available.}}
\email{renzhi@westlake.edu.cn}

\affiliation{$^{1}$Department of Physics, School of Science, Westlake University, 18 Shilongshan Road, Hangzhou, 310024, Zhejiang Province, PR China}
\affiliation{$^{2}$Institute of Natural Sciences, Westlake Institute for Advanced Study, 18 Shilongshan Road, Hangzhou, 310024, Zhejiang Province, PR China}
\affiliation{$^{3}$Department of Physics, Fudan University, Shanghai 200433, PR China}

\begin{abstract}
The recent claim of potential room-temperature superconductivity in Pb$_{10-x}$Cu$_{x}$(PO$_{4}$)$_{6}$O has attracted widespread attention. However, the signature of superconductivity is later attributed to the Cu$_{2}$S impurity formed during the multiple-step synthesis procedure. Here we report a simple one-step approach to synthesize single-phase chloride analogue Cu-doped Pb$_{10}$(PO$_{4}$)$_{6}$Cl$_{2}$ using PbO, PbCl$_{2}$, CuCl$_{2}$, and NH$_{4}$H$_{2}$PO$_{4}$ as starting materials.
Irrespective of the initial stoichiometry, the Cu doping always leads to a lattice expansion in Pb$_{10}$(PO$_{4}$)$_{6}$Cl$_{2}$.
This indicates that Cu prefers to reside in the hexagonal channels rather than substitutes at the Pb site, and the chemical formula is expressed as Pb$_{10}$(PO$_{4}$)$_{6}$Cu$_{x}$Cl$_{2}$.
All the Pb$_{10}$(PO$_{4}$)$_{6}$Cu$_{x}$Cl$_{2}$ (0 $\leq$ $x$ $\leq$ 1.0) samples are found to be semiconductors with wide band gaps of 4.46-4.59 eV, and the Cu-doped ones ($x$ = 0.5 and 1.0) exhibit a paramagnetic behavior without any phase transition between 400 and 1.8 K.
Our study calls for a reinvestigation of the Cu location in Pb$_{10-x}$Cu$_{x}$(PO$_{4}$)$_{6}$O, and supports the absence of superconductivity in this oxyapatite.
\end{abstract}

\maketitle
\maketitle

\noindent\textbf{1. Introduction}\\

Achieving room temperature superconductivity has been the dream in the field of superconducting material research since it may bring revolutionary applications in diverse fields, such as magnetic leviation, power transmission and electronic device \cite{RTSC1,RTSC2,RTSC3}.
In this context, the report of potential superconductivity at 378 K in Cu-doped lead apatite Pb$_{10-x}$Cu$_{x}$(PO$_{4}$)$_{6}$O (0.9 $<$ $x$ $<$ 1.1) (so called LK99) at ambient condition \cite{LK991,LK992} has generated intense attention worldwide.
However, this observation cannot be reproduced despite many attempts in subsequent studies \cite{repeat1,repeat2,repeat3,repeat4,repeat5,repeat6,repeat7,repeat8,repeat9,repeat10}.
Instead, a very recent study suggests that the superconducting-like resistivity drop in LK99 is most likely caused by the first-order transition of the Cu$_{2}$S impurity, which is formed between the reaction of Cu$_{3}$P and Pb$_{2}$(SO$_{4}$)O in the multiple-step synthesis process \cite{repeat8,repeat10}.

The parent compound Pb$_{10}$(PO$_{4}$)$_{6}$O belongs to the Pb$_{10}$(PO$_{4}$)$_{6}$$X_{2}$ apatite family, where $X$ can be, in addition to O, F, Cl, Br, I, and OH \cite{apatite}.
The crystal structure of this family can be regarded as an anion-stuffed derivative of the Mn$_{5}$Si$_{3}$-type ($D$8$_{8}$ type) and adopts the hexagonal $P$6$_{3}$/$m$ space group.
Especially, $X$ ions occupy the central axes of the hexagonal channels formed by Pb$^{2+}$ ones [see Fig. 1(a)], and, in the case of O, the charge balance is maintained through the creation of O vacancies.
While Cu is supposed to substitute for Pb in Pb$_{10}$(PO$_{4}$)$_{6}$O \cite{LK991,LK992}, it is found located in the hexagonal channels of other Cu-containing apatites such as Sr$_{5}$(VO$_{4}$)$_{3}$CuO \cite{SrVCuO}.
This naturally raises the question about the location of Cu and its doping effect in other Pb$_{10}$(PO$_{4}$)$_{6}$$X_{2}$ apatites.

\begin{table*}
	\caption{Preparation condition, structural and chemical characterization of the pristine and Cu-doped Pb$_{10}$(PO$_{4}$)$_{6}$Cl$_{2}$ samples.}
	\renewcommand\arraystretch{1.2}
	\begin{tabular}{p{3.3cm}<{\centering}p{4.0cm}<{\centering}p{2.4cm}<{\centering}p{1.5cm}<{\centering}p{1.5cm}<{\centering}p{1.5cm}<{\centering}p{2.5cm}<{\centering}}
		\\
		\hline
		Nominal composition       & Synthesis temperature ($^{\circ}$C) & Phase purity & $a$ ({\AA}) &  $c$ ({\AA}) &  $V$ ({\AA}$^{3}$) & Measured Cu/Pb\\
		\hline
		Pb$_{10}$(PO$_{4}$)$_{6}$Cl$_{2}$&880&single &9.967(1)&7.336(1)&631.1&$-$\\
        Pb$_{10}$(PO$_{4}$)$_{6}$Cu$_{0.5}$Cl$_{2}$ &  850 &single &9.975(1)&7.333(1)&631.9&0.03(1)\\
		Pb$_{10}$(PO$_{4}$)$_{6}$CuCl$_{2}$  & 800&single&9.980(1)&7.333(1)&632.5&0.09(1)\\
        Pb$_{9.5}$Cu$_{0.5}$(PO$_{4}$)$_{6}$Cl$_{2}$ & 850&single&9.973(1)&7.333(1)&631.4&$-$\\
        Pb$_{9}$Cu(PO$_{4}$)$_{6}$Cl$_{2}$ &800&single&9.975(1)&7.330(1)&631.6&$-$\\
		\hline
	\end{tabular}
	\label{Table1}
\end{table*}

In this paper, we present the synthesis and properties of pristine and Cu-doped Pb$_{10}$(PO$_{4}$)$_{6}$Cl$_{2}$ apatites.
For the first time, single-phase samples were obtained by a one-step method.
The location of Cu is investigated by comparing two different Pb$_{10-x}$Cu$_{x}$(PO$_{4}$)$_{6}$Cl$_{2}$ and Pb$_{10}$(PO$_{4}$)$_{6}$Cu$_{x}$Cl$_{2}$ series.
In the latter case, the Cu/
Pb binding energies, optical absorption spectra, and magnetic properties are characterized.
The implications of these results are also discussed.\\

\noindent\textbf{2. Experimental section}\\

\emph{Synthesis of Pb$_{10}$(PO$_{4}$)$_{6}$Cl$_{2}$}. High purity powders of PbO (99.9\%), PbCl$_{2}$ (99.9\%), and NH$_{4}$H$_{2}$PO$_{4}$ (99.9\%) were used as starting materials and weighed according to reaction (1).
\begin{equation}
\begin{split}
\rm 9PbO+PbCl_{2}+6NH_{4}H_{2}PO_{4}\rightarrow Pb_{10}(PO_{4})_{6}Cl_{2}+ \\ \rm 6NH_{3}\uparrow
+9H_{2}O\uparrow
\end{split}
\end{equation}
The powders were mixed thoroughly using a mortar and pestle and pressured into a pellet in an argon filled glove box. The pellet was then heated in air at 880 $^{\circ}$C for 24 hours, followed by furnace cooling to room temperature.
Since PbCl$_{2}$ starts to evaporate above $\sim$270 $^{\circ}$C \cite{PbCl2}, samples with 10\% and 20\% excess of PbCl$_{2}$ were also prepared by the same procedure to compensate possible loss during the heating process.

\emph{Synthesis of Cu-doped Pb$_{10}$(PO$_{4}$)$_{6}$Cl$_{2}$}.
Two series of Cu-doped Pb$_{10}$(PO$_{4}$)$_{6}$Cl$_{2}$ apatite samples were prepared and CuCl$_{2}$ (99.99\%) was employed as the Cu source.
For the Pb$_{10-x}$Cu$_{x}$(PO$_{4}$)$_{6}$Cl$_{2}$ series, where Cu was assumed to substitute for Pb, the powders of PbO, PbCl$_{2}$, NH$_{4}$H$_{2}$PO$_{4}$ and CuCl$_{2}$ were weighed according to reaction (2).
\begin{equation}
\begin{split}
\rm 9PbO+(1-\emph{x})PbCl_{2}+\emph{x}CuCl_{2}+6NH_{4}H_{2}PO_{4}\rightarrow \\
\rm Pb_{10-\emph{x}}Cu_{\emph{x}}(PO_{4})_{6}Cl_{2}+6NH_{3}\uparrow+\rm 9H_{2}O\uparrow
\end{split}
\end{equation}
For the Pb$_{10}$(PO$_{4}$)$_{6}$Cu$_{x}$Cl$_{2}$ series, where Cu was assumed to occupy the hexagonal tunnel position, the powders of PbO, PbCl$_{2}$, NH$_{4}$H$_{2}$PO$_{4}$ and CuCl$_{2}$ were weighed according to reaction (3).
\begin{equation}
\begin{split}
\rm (9+\emph{x})PbO+(1-\emph{x})PbCl_{2}+\emph{x}CuCl_{2}+6NH_{4}H_{2}PO_{4}\rightarrow \\
\rm Pb_{10}(PO_{4})_{6}Cu_{\emph{x}}Cl_{2}+(6-2\emph{x}/3)NH_{3}\uparrow+(\emph{x}/3)N_{2}\uparrow \\
+\rm (9+\emph{x})H_{2}O\uparrow
\end{split}
\end{equation}
In both cases, Cu content $x$ was set to 0.5 and 1.0. The powders were mixed thoroughly using a mortar and pestle and pressured into pellets in an argon filled glove box.
The pellets were then heated in air at 800$-$850 $^{\circ}$C for 24 hours, followed by furnace cooling to room temperature.
The synthesis temperatures, which depend on the Cu content $x$, are listed in Table I.
\begin{figure*}
	\includegraphics*[width=16.8cm]{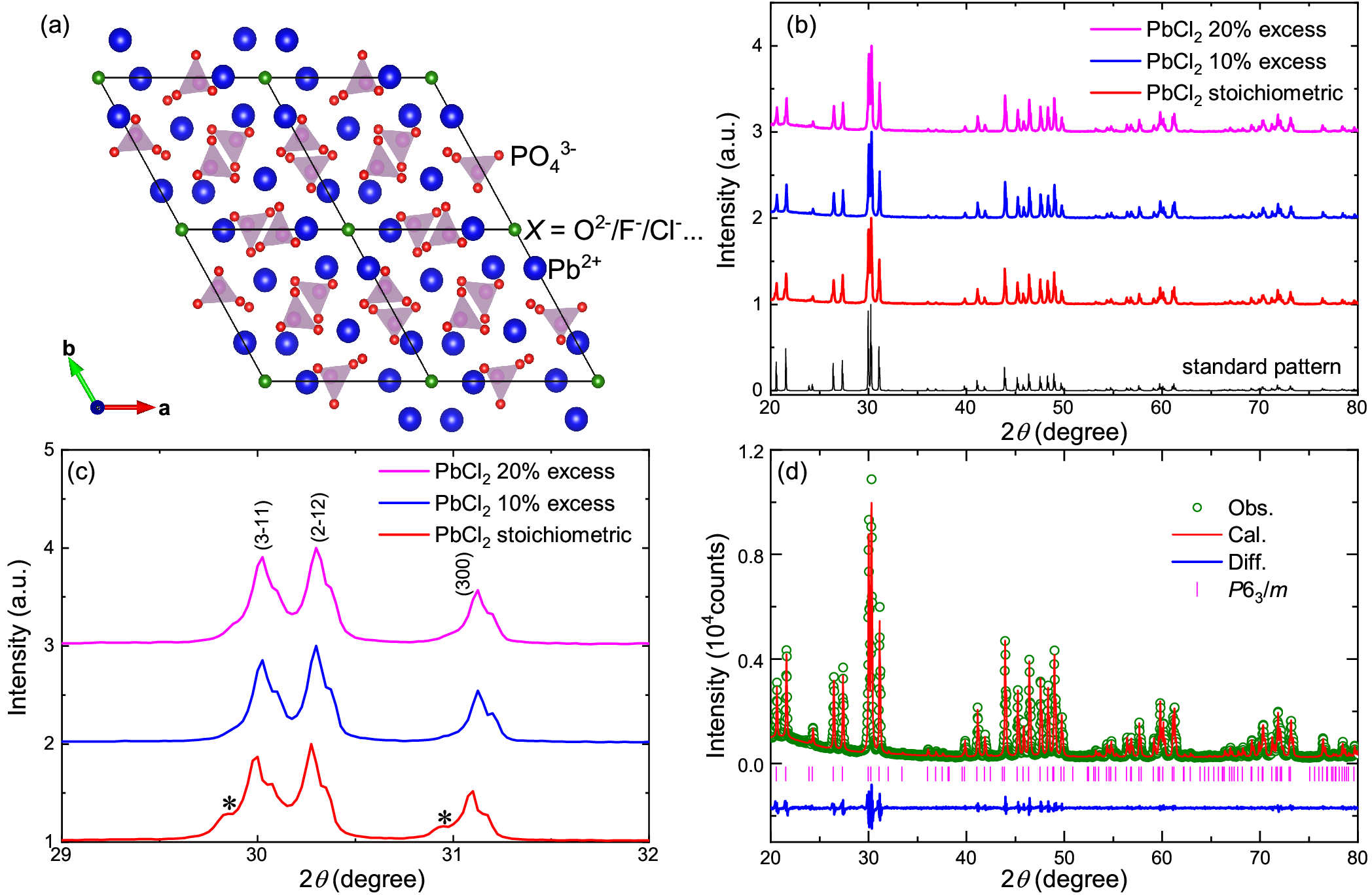}
	\caption{
			(a) Schematic structure of the Pb$_{10}$(PO$_{4}$)$_{6}$$X$$_{2}$ apatite, in which the Pb$^{2+}$, PO$_{4}^{3-}$ and $X$ ions are labeled.
            (b) XRD patterns for the Pb$_{10}$(PO$_{4}$)$_{6}$Cl$_{2}$ apatite samples prepared using different amounts of PbCl$_{2}$ in starting material. The standard pattern is also included for comparison.
            (c) A zoom of the patterns in the 2$\theta$ range of 29$^{\circ}$ to 32$^{\circ}$. The diffraction peaks from the apatite phase are indexed and impurity peaks are labeled by the asterisks.
            (d) Structural refinement profile of the phase pure sample using the Le Bail method.}
	\label{fig2}
\end{figure*}
\begin{figure*}
	\includegraphics*[width=16.6cm]{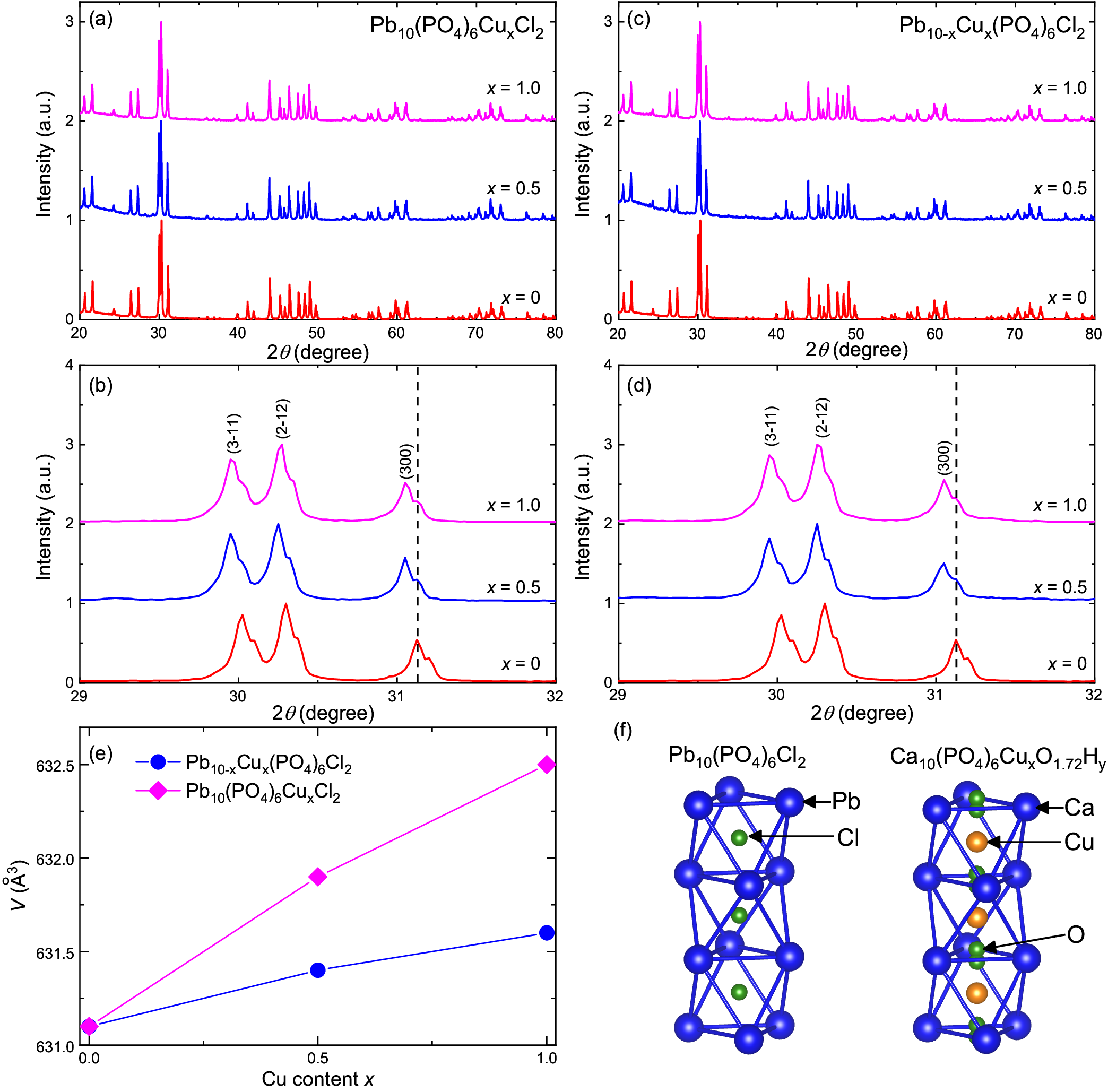}
	\caption{
			(a) XRD patterns for the series of Pb$_{10}$(PO$_{4}$)$_{6}$Cu$_{x}$Cl$_{2}$ samples. (b) A zoom of the patterns in the 2$\theta$ range of 29$^{\circ}$ to 32$^{\circ}$.
               The diffraction peaks are indexed and the vertical line is a guide to the eyes.
            (c,d) Same set of data for the series of Pb$_{10-x}$Cu$_{x}$(PO$_{4}$)$_{6}$Cl$_{2}$ samples.
            (e) Cu content $x$ dependencies of the unit-cell volume.
            (f) Schematic structure of the hexagonal tunnels of Pb$_{10}$(PO$_{4}$)$_{6}$Cl$_{2}$ and Ca$_{5}$(PO$_{4}$)$_{3}$Cu$_{x}$O$_{0.86}$H$_{y}$. The different atoms are labeled by the arrows.
	}
	\label{fig2}
\end{figure*}

\emph{Sample characterization}.
The phase purity of the resulting samples was checked by powder x-ray diffraction (XRD) at room temperature using a Bruker D8 Advance X-ray diffractometer with Cu K$\alpha$ radiation.
The structural refinements were performed using the JANA2006 software.
The morphology and chemical composition were characterized with a scanning electron microscope (SEM, Hitachi) equipped with an energy-dispersive x-ray (EDX) spectrometer.
The XPS spectra were recorded in an ESCALAB Xi+ spectrometer with Al K$\alpha$ x-rays as the excitation source.
The optical adsorption spectra were acquired using a UV-vis spectrophotometer in the range of 200-500 nm.
The dc magnetization measurements were done in a commercial SQUID magnetometer (MPMS3).\\

\noindent\textbf{3. Results and Discussion}\\

\noindent \emph{3.1. X-ray diffraction}\\

Figure 1(b) shows the XRD patterns for the Pb$_{10}$(PO$_{4}$)$_{6}$Cl$_{2}$ samples prepared with different amounts of PbCl$_{2}$.
At fist glance, all diffraction patterns look similar and agree well with the standard apatite phase (the bottom one).
Nevertheless, small impurity peaks are visible at 2$\theta$ $\approx$ 29.8$^{\circ}$ and 30.5$^{\circ}$ on zooming the patterns [see Fig. 1(b)] for the sample prepared using stoichiometric PbCl$_{2}$ (marked by the asterisks).
It is thus clear that PbCl$_{2}$ evaporates during the heating process and its excess is necessary to eliminate the impurities.
In the unit cell of Pb$_{10}$(PO$_{4}$)$_{6}$Cl$_{2}$, there are two Pb sites, one P site, three O sites, and one Cl site, as listed in Table II.
Based on this structural model, the structural refinement is performed for the sample with 10\% excess of PbCl$_{2}$ and the result is displayed in Fig. 1(d).
The calculated pattern (solid line) matches well the observed one (open circles) with small reliability factors of $R_{\rm wp}$ = 6.5\% and $R_{\rm p}$ = 4.8\%, confirming the single-phase nature.
The refined lattice parameters are $a$ = 9.967(1) {\AA} and $c$ = 7.336(1) {\AA}, in good agreement with those reported previously \cite{apatite}.
\begin{table}[b]
	\caption{Atomic coordinates for the structural refinement of Pb$_{10}$(PO$_{4}$)$_{6}$Cl$_{2}$.}
	\renewcommand\arraystretch{1.2}
	\begin{tabular}{p{2cm}<{\centering}p{2cm}<{\centering}p{2cm}<{\centering}p{2cm}<{\centering}}
		\\
		\hline
		Atom         & $x$ & $y$ & $z$\\
		\hline
		Pb1& 0.3333& 0.6667& 0.0051 \\
        Pb2& 0.2547& 0.2478& 0.25 \\
        P& 0.4102& 0.0312 & 0.25\\
        O1 & 0.3402& -0.1491& 0.25\\
        O2 & 0.5871& 0.1136& 0.25\\
        O3 & 0.3593& 0.0855& 0.0848\\
        Cl & 0& 0& 0\\
		
		\hline
	\end{tabular}
	\label{Table1}
\end{table}

\begin{figure}
	\includegraphics*[width=7.3cm]{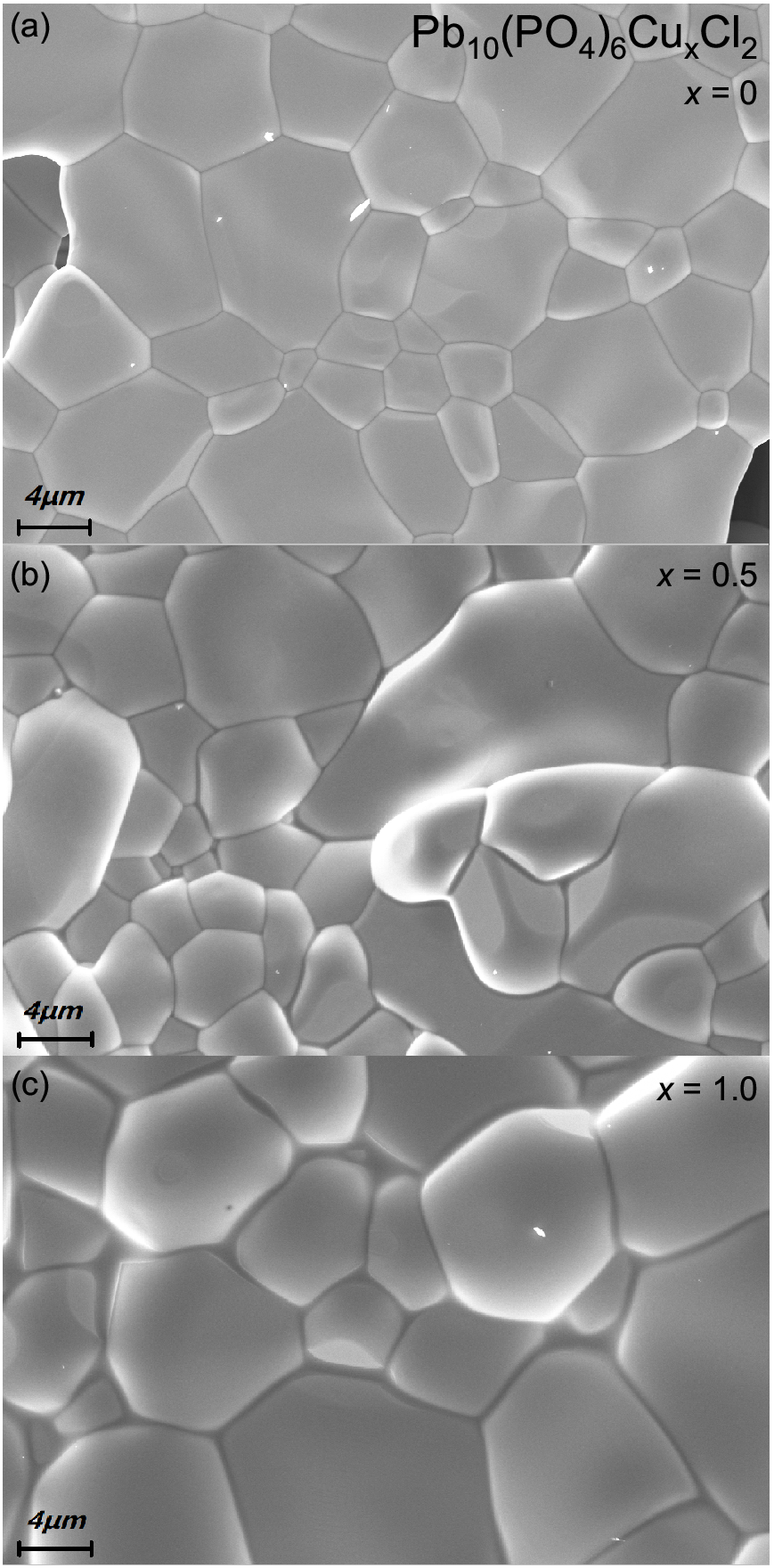}
	\caption{
	(a-c) SEM images with a scale bar of 4 $\mu$m for the Pb$_{10}$(PO$_{4}$)$_{6}$Cu$_{x}$Cl$_{2}$ samples with $x$ = 0, 0.5, and 1.0, respectively.
	}
	\label{fig3}
\end{figure}
The XRD patterns for the Cu-doped Pb$_{10}$(PO$_{4}$)$_{6}$Cl$_{2}$ samples are shown in Fig. 2.
For both the Pb$_{10}$(PO$_{4}$)$_{6}$
Cu$_{x}$Cl$_{2}$ and Pb$_{10-x}$Cu$_{x}$(PO$_{4}$)$_{6}$Cl$_{2}$ series, a single apatite phase is retained up to $x$ = 1.
As seen in the zooms of the patterns in Figs. 2(b) and (d), the diffraction peaks shift towards lower 2$\theta$ values.
By a least-squares fitting method, the lattice parameters are determined and listed in Table I.
Surprisingly, a similar trend is observed for both Pb$_{10}$(PO$_{4}$)$_{6}$Cu$_{x}$Cl$_{2}$ and Pb$_{10-x}$Cu$_{x}$(PO$_{4}$)$_{6}$Cl$_{2}$ series.
Namely, with increasing $x$, the $a$-axis increases monotonically while the $c$-axis remains almost unchanged, resulting in a lattice expansion shown in Fig. 2(e).
Note that the ionic radii are 1.20 {\AA}, 0.96 {\AA}, 0.72 {\AA}, and 1.81 {\AA} for Pb$^{2+}$, Cu$^{+}$, Cu$^{2+}$ and Cl$^{-}$, respectively.
Hence the substitution of Pb by Cu is expected to shrink the lattice, which is at odds with the experimental observation.
It should be pointed out that Cu was found to reside in the hexagonal channels in many Cu-containing apatites such as Sr$_{5}$(VO$_{4}$)$_{3}$CuO \cite{SrVCuO} and Ca$_{5}$(PO$_{4}$)$_{3}$Cu$_{x}$O$_{0.86}$H$_{y}$ \cite{Cu-contaning apatite}.
In those cases, the unit-cell volume expands upon Cu doping, as observed in the present study.
Taken together, these provide supporting evidence that Cu also resides in the hexagonal channels of Cu-doped Pb$_{10}$(PO$_{4}$)$_{6}$Cl$_{2}$.
Note that, in Ca$_{5}$(PO$_{4}$)$_{3}$Cu$_{x}$O$_{0.86}$H$_{y}$ \cite{Cu-contaning apatite}, the Cu takes the center [(0,0,0) site] of the Ca$^{2+}$ octahedra while O is located near the center [(0,0,$\sim$0.2) site] of the planar Ca$^{2+}$ triangles [see Fig. 2(f)].
In the case of Cu-doped Pb$_{10}$(PO$_{4}$)$_{6}$Cl$_{2}$, however, the center of Pb$^{2+}$ octahedra is already occupied by the Cl$^{-}$ ions.
Hence the exact atomic coordinates of Cu remain unclear and requires single crystal diffraction studies in future. In the remaining part of the paper, we focus on the series of Pb$_{10}$(PO$_{4}$)$_{6}$Cu$_{x}$Cl$_{2}$ samples.\\

\noindent \emph{3.2. SEM and EDX characterization}\\

Figures 3(a-c) show the SEM images with a scale bar of 4 $\mu$m for the Pb$_{10}$(PO$_{4}$)$_{6}$Cu$_{x}$Cl$_{2}$ samples with $x$ = 0, 0.5 and 1, respectively. All samples are homogeneous and free from impurities with a granular microstructure.
The grain size is comparable and up to $\sim$10 $\mu$m for $x$ = 0 and 0.5, while is increased up to a few tenth $\mu$m for $x$ = 1.
Since the latter sample is prepared at a lower temperature (see Table I), it is reasonable to speculate that the grain growth is facilitated by the the addition of CuCl$_{2}$. On the other hand, the EDX measurements confirm the presence of Cu in both Cu-doped samples and give the Cu:Pb ratios of 0.03(1) and 0.09(1) for $x$ = 0.5 and 1, respectively. These values are in reasonable agreement with the nominal compositions within the experimental uncertainty. In passing, the EDX elemental mapping results indicate that all the constituent elements are uniformly distributed for these samples (see ESI Fig. 1).\\

\noindent \emph{3.3. XPS spectra}\\
\begin{figure}
	\includegraphics*[width=8.6cm]{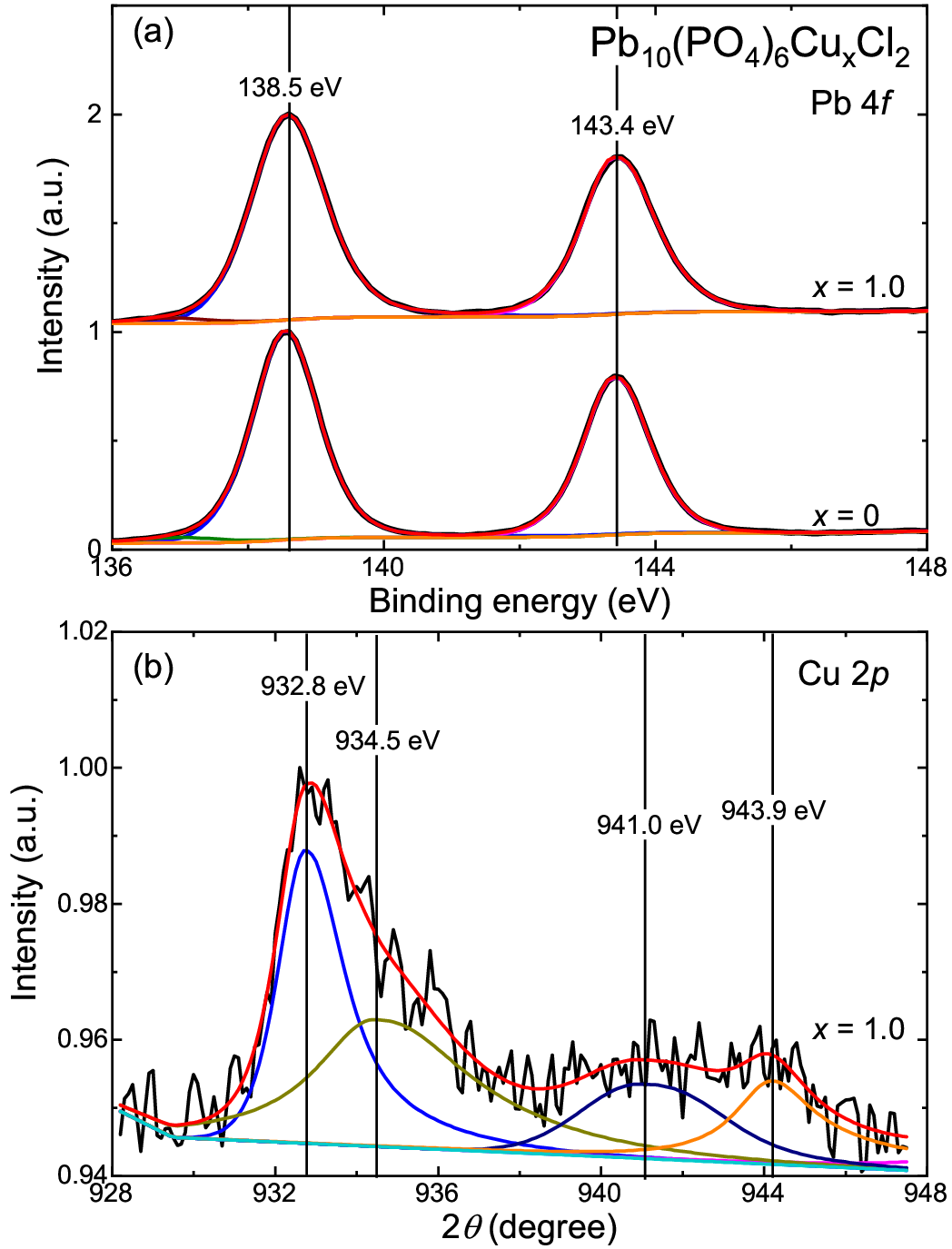}
	\caption{
	(a) XPS spectra and corresponding deconvolutions in the Pb 4$f$ region for the Pb$_{10}$(PO$_{4}$)$_{6}$Cu$_{x}$Cl$_{2}$ samples with $x$ = 0 and 1.0. The vertical lines are guides to the eyes, showing the binding energies.
    (b) XPS spectrum and corresponding deconvolution in the Cu 2$p$ region for the sample with $x$ = 1.0. The vertical lines are guides to the eyes, showing the binding energies.
	}
	\label{fig3}
\end{figure}

\begin{figure}
	\includegraphics*[width=8.7cm]{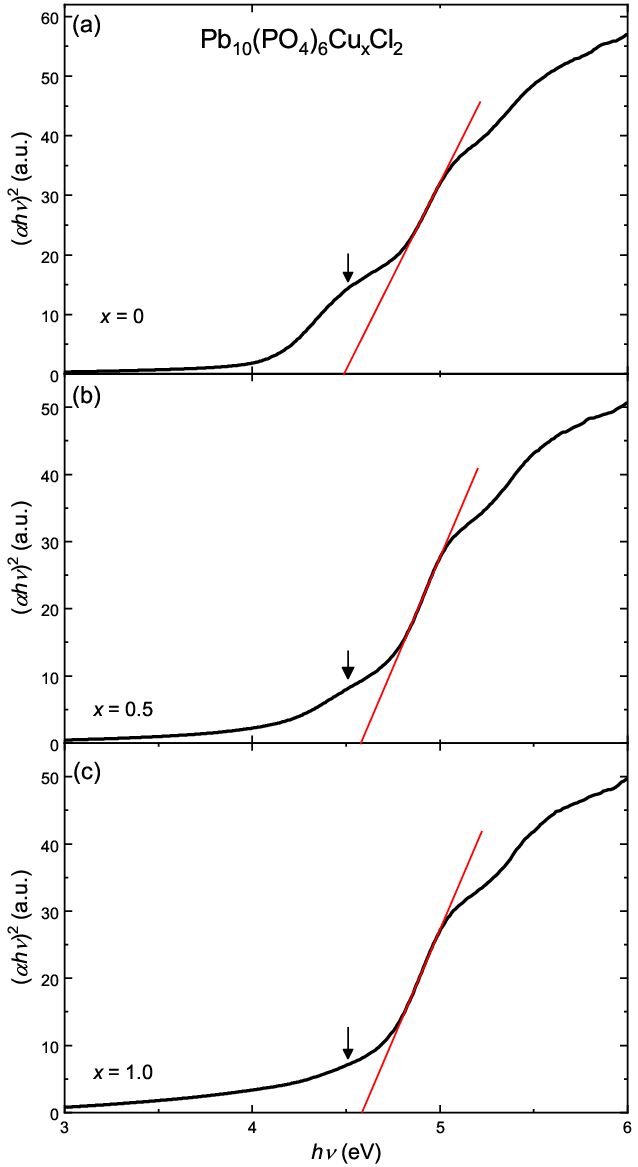}
	\caption{
	 (a-c) Photo energy dependence of UV-Vis absorbance spectra for the Pb$_{10}$(PO$_{4}$)$_{6}$Cu$_{x}$Cl$_{2}$ samples with $x$ = 0, 0.5, and 1.0, respectively.
      In each panel, the line denotes the extrapolation of the curve to determine the band gap and the arrow mark the shoulder below the steepest slope.
	}
	\label{fig5}
\end{figure}

The XPS spectra in the Pb 4$f$ region for the Pb$_{10}$(PO$_{4}$)$_{6}$Cu$_{x}$Cl$_{2}$ samples with $x$ = 0 and 1 are displayed in Fig. 4(a).
For both samples, the deconvolution of the spectrum gives a pair of binding energies located at 138.5 eV and 143.4 eV, which agree well with those found in PbO \cite{PbXPS}. Nevertheless, the peak width is broader for $x$ = 1 than for $x$ = 0, pointing to increased structural disorder resulting from Cu doping. In the Cu 2$p$ region, only the spectrum for $x$ = 1.0 can be resolved, as shown in Fig. 4(b). The deconvolution of the spectrum gives four binding energies located at 932.8 eV, 934.5 eV, 941.0 eV and 943.9 eV. The former two binding energies correspond well to the Cu$^{+}$ and Cu$^{2+}$, respectively, while the latter two ones are typical satellites for Cu$^{2+}$ spices \cite{CuXPS}.
According to a previous study \cite{Custate}, the oxidation state of Cu in the hexagonal channels of Sr$_{5}$(PO$_{4}$)$_{3}$Cu$_{x}$OH$_{y}$ can be varied by annealing in different atmospheres. All copper is oxidized to Cu$^{2+}$ upon annealing in oxygen while reduced to Cu$^{+}$ during annealing with a CuO/Cu$_{2}$O getter. Since our samples were prepared in air, a mixture of Cu$^{+}$ and Cu$^{2+}$ is reasonable \cite{Cu-contaning apatite}, and the charge balance may be maintained by either simultaneous incorporation of negative Cl$^{-}$ ions in the channels or the formation of impurity levels. In fact, we have also measured the Cl content by the EDX method, which indicates that the ratio of Cu:Cl is close to $x$:2. Nevertheless, given the small atomic number of Cl, this result should be considered as semiquantitative. Due to the large difference in atomic weight between Cu and Cl, it is almost impossible to determine their contents accurately by the same analysis technique. Nevertheless, it is unlikely that all Cu exists in the form of CuCl$_{2}$ in the hexagonal channels considering its mixed valence states. \\

\noindent \emph{3.4. UV-vis spectroscopy}\\

The UV-vis spectra for the Pb$_{10}$(PO$_{4}$)$_{6}$Cu$_{x}$Cl$_{2}$ samples with 0 $\leq$ $x$ $\leq$ 1.0 are plotted as $(\alpha h\nu)$$^{2}$ versus $h\nu$ in Figs. 5(a-c).
Here $\alpha$ is the absorbance coefficient and $h\nu$ is the photon energy.
Following the Tauc method \cite{Tauc}, the direct energy gap $E_{\rm g}$ is determined by the intercept between the extrapolation of the steepest slop and the photon-energy axis, as indicated by the red lines.
This gives $E_{\rm g}$ = 4.46 eV, 4.57 eV, and 4.59 eV for $x$ = 0, 0.5, and 1.0, respectively, which are comparable to the reported values for isostructural apatites such as Ca$_{10}$(PO$_{4}$)$_{6}$Cl$_{2}$ ($\sim$5.3 eV) \cite{bandgap}.
Whereas $E_{\rm g}$ remains almost unchanged, it is seen that the shoulder below the steepest $(\alpha h\nu)$$^{2}$ slope (see the arrows) is gradually diminished with increasing Cu content $x$.
Thus the Cu doping induces a significant change in the indirect optical transition, possibly related to the in-gap impurity states. \\
\begin{figure}
	\includegraphics*[width=8.7cm]{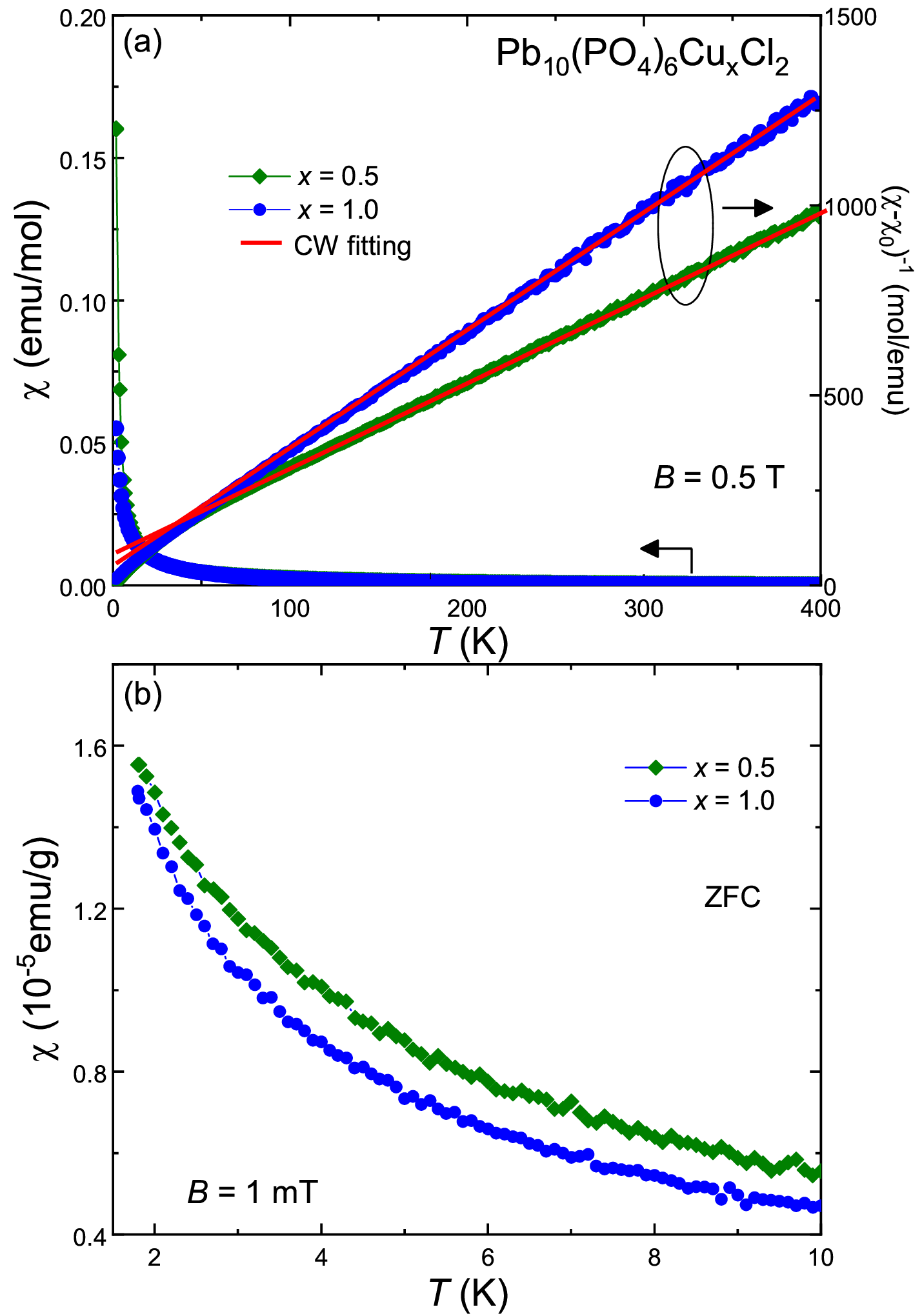}
	\caption{
			(a) Left axis: Temperature dependencies of magnetic susceptibility measured under an applied field of 0.5 T for the Pb$_{10}$(PO$_{4}$)$_{6}$Cu$_{x}$Cl$_{2}$ samples with $x$ = 0.5 and 1.0. Right axis: Temperature dependencies of the inverse magnetic susceptibility. The solid lines are fits to the data by the Curie-Weiss model.
            (b) Low-temperature magnetic susceptibility measured under an applied field of 1 mT for the same samples.
	}
	\label{fig2}
\end{figure}

\noindent \emph{3.5. Magnetic susceptibility}\\

Figure 6(a) shows the temperature dependence of the magnetic susceptibility for the Pb$_{10}$(PO$_{4}$)$_{6}$Cu$_{x}$Cl$_{2}$ samples with $x$ = 0.5 and 1.0 measured between 400 and 1.8 K under an applied field of 0.5 T.
For both samples, the susceptibility displays a paramagnetic behavior and can be well fitted by the Curie-Weiss law as follows,
\begin{equation}\label{1}
  \chi =  \chi_{0} + \frac{C}{T-\Theta},
\end{equation}
where $\chi_{0}$ is the temperature-independent term, $C$ is the Curie constant and $\Theta$ is the Weiss temperature.
This is illustrated by the plot of inverse susceptibility and the best fits yield $\chi_{0}$ = $-$6.05$\times$10$^{-4}$ emu/mol, $C$ = 0.318 emuK/mol, $\Theta$ = $-$12 K for $x$ = 0.5, and $\chi_{0}$ = $-$5.97$\times$10$^{-4}$ emu/mol, $C$ = 0.445 emuK/mol, $\Theta$ = $-$38 K for $x$ = 1.0.
The negative $\Theta$ indicates an antiferromagnetic interaction between the local moments.
Assuming $C$ is due to the Cu$^{2+}$ ($S$ = 1/2), the effect magnetic moment $\mu_{\rm eff}$ is calculated to be 2.26 $\mu_{\rm B}$ and 1.87 $\mu_{\rm B}$ for $x$ = 0.5 and 1.0, respectively, where $\mu_{\rm B}$ is the Bohr magneton. Both $\mu_{\rm eff}$ values are larger than the spin-1/2 value of 1.73 $\mu_{\rm B}$ but typically observed for power samples containing Cu$^{2+}$ ions \cite{Cu2+ions}. In addition, the smaller $\mu_{\rm eff}$ for $x$ = 1.0 compared with that for $x$ = 0.5 is consistent with the presence of nonmagnetic Cu$^{+}$ ions in the former case.
As shown in Fig. 6(b), the zero-field cooling susceptibility curve still displays a paramagnetic behavior even if the applied magnetic field is reduced to 1 mT.
Obviously, the Cu-doped Pb$_{10}$(PO$_{4}$)$_{6}$Cl$_{2}$ samples show neither superconductivity nor any other phase transition between 400 and 1.8 K.\\

\noindent \emph{3.6. Implication on the study of Cu-doped Pb$_{10}$(PO$_{4}$)$_{6}$O}\\

Finally, we briefly discuss the implication of our results on the study of Cu-doped Pb$_{10}$(PO$_{4}$)$_{6}$O.
First, we offer a simple one-step method to synthesis Cu-doped Pb$_{10}$(PO$_{4}$)$_{6}$Cl$_{2}$ apatites.
By replacing the chlorides with the oxide counterparts in the starting materials, such approach should also be applicable to the Cu-doped Pb$_{10}$(PO$_{4}$)$_{6}$O system.
This avoids the introduction of foreign elements and allows for a better evaluation of the intrinsic properties.
Second, our results suggest a portion of Cu may also occupy the central axes of the hexagonal tunnels in Cu-doped Pb$_{10}$(PO$_{4}$)$_{6}$O, in particular given that half of the O sites in the tunnels are vacant.
Further studies are needed to clarify this issue and examine its effect on the properties.
Third, our results show unambiguously that Cu-doped Pb$_{10}$(PO$_{4}$)$_{6}$Cl$_{2}$ exhibits paramagnetic semiconducting behavior below 400 K.
This supports other studies \cite{repeat1,repeat2,repeat3,repeat4,repeat5,repeat6,repeat7,repeat8,repeat9,repeat10} that indicate the absence of room-temperature superconductivity in Cu-doped Pb$_{10}$(PO$_{4}$)$_{6}$O.\\

\noindent\textbf{4. Conclusion}\\

In summary, motivated by the recent report of potential room-temperature superconductivity in Pb$_{10-x}$Cu$_{x}$(PO$_{4}$)$_{6}$O, we have explored the synthesis and properties of pristine and Cu-doped Pb$_{10}$(PO$_{4}$)$_{6}$Cl$_{2}$.
Using PbO, PbCl$_{2}$, CuCl$_{2}$, and NH$_{4}$H$_{2}$PO$_{4}$ as starting materials, single-phase samples can be prepared by a simple one-step method.
Irrespective of the initial stoichiometry, the Cu doping always leads to an expansion of the hexagonal lattice in Pb$_{10}$(PO$_{4}$)$_{6}$Cl$_{2}$.
Thus Cu prefers to reside in the hexagonal channels rather than substitutes at the Pb site, giving the chemical formula Pb$_{10}$(PO$_{4}$)$_{6}$Cu$_{x}$Cl$_{2}$.
All the Pb$_{10}$(PO$_{4}$)$_{6}$Cu$_{x}$Cl$_{2}$ (0 $\leq$ $x$ $\leq$ 1.0) samples turn out to be wide-gap semiconductors, and a paramagnetic behavior without any phase transition between 400 and 1.8 K is observed for both $x$ = 0.5 and 1.0.
Our study calls for a reinvestigation of the Cu location and the ground state properties in Pb$_{10-x}$Cu$_{x}$(PO$_{4}$)$_{6}$O.

\section*{ACKNOWLEDGEMENT}
We acknowledge financial support by the foundation of Westlake University and technical support from the Instrumentation and Service Center for Molecular Sciences at Westlake University for SEM, XPS and UV-vis measurements.

\end{document}